%% file: arxiv.tex
\documentclass[10pt,amsmath,amssymb,aps,physrev,superscriptaddress,showkeys]{revtex4-2}

\usepackage[resetlabels,labeled]{multibib}
\newcites{S}{References}

\usepackage{graphicx}
\usepackage{dcolumn}
\usepackage{bm}
\usepackage[T1]{fontenc}
\usepackage{siunitx}
\DeclareSIUnit\sq{sq}
\DeclareSIUnit\OSq{\ohm \per \sq}
\DeclareSIUnit\Torr{Torr}
\DeclareSIUnit\mt{\milli\Torr}

\usepackage{amsmath}
\usepackage[colorlinks=true, allcolors=blue]{hyperref}

\newcommand{\fr}{\text{Fr}}

\begin{document}

\title{Large-Area Metal-Integrated Grating Electrode Achieving Near 100\% Infrared Transmission}

\author{Karolina Bogdanowicz}
\affiliation{Łukasiewicz Research Network -- Institute of Microelectronics and Photonics, al.~Lotników 32/46, 02-668 Warsaw, Poland}
\affiliation{Photonics Group, Institute of Physics, Lodz University of Technology, ul.~Wólczańska 219, 90-924 Łódź, Poland}
\author{Weronika Głowadzka}
\affiliation{Łukasiewicz Research Network -- Institute of Microelectronics and Photonics, al.~Lotników 32/46, 02-668 Warsaw, Poland}
\affiliation{Photonics Group, Institute of Physics, Lodz University of Technology, ul.~Wólczańska 219, 90-924 Łódź, Poland}
\author{Tristan Smołka}
\affiliation{Laboratory for Optical Spectroscopy of Nanostructures, Department of Experimental Physics, Faculty of Fundamental Problems of Technology, Wrocław University of Science and Technology, Wybrzeże Wyspiańskiego 27, 50-370 Wrocław, Poland}
\author{Michał Rygała}
\affiliation{Laboratory for Optical Spectroscopy of Nanostructures, Department of Experimental Physics, Faculty of Fundamental Problems of Technology, Wrocław University of Science and Technology, Wybrzeże Wyspiańskiego 27, 50-370 Wrocław, Poland}
\author{Marcin Kałuża}
\affiliation{Institute of Electronics, Lodz University of Technology, al.~Politechniki 8, 93-590 Łódź, Poland}
\author{Marek Ekielski}
\affiliation{Łukasiewicz Research Network -- Institute of Microelectronics and Photonics, al.~Lotników 32/46, 02-668 Warsaw, Poland}
\author{Oskar Sadowski}
\affiliation{Łukasiewicz Research Network -- Institute of Microelectronics and Photonics, al.~Lotników 32/46, 02-668 Warsaw, Poland}
\affiliation{Warsaw University of Technology -- Institute of Microelectronics and Optoelectronics, 00-662 Warsaw, Koszykowa 75, Poland}
\author{Magdalena Zadura}
\affiliation{Łukasiewicz Research Network -- Institute of Microelectronics and Photonics, al.~Lotników 32/46, 02-668 Warsaw, Poland}
\affiliation{Photonics Group, Institute of Physics, Lodz University of Technology, ul.~Wólczańska 219, 90-924 Łódź, Poland}
\author{Magdalena Marciniak}
\affiliation{Photonics Group, Institute of Physics, Lodz University of Technology, ul.~Wólczańska 219, 90-924 Łódź, Poland}
\author{Marcin Gębski}
\affiliation{Photonics Group, Institute of Physics, Lodz University of Technology, ul.~Wólczańska 219, 90-924 Łódź, Poland}
\author{Michał Wasiak}
\affiliation{Photonics Group, Institute of Physics, Lodz University of Technology, ul.~Wólczańska 219, 90-924 Łódź, Poland}
\author{Marcin Motyka}
\affiliation{Laboratory for Optical Spectroscopy of Nanostructures, Department of Experimental Physics, Faculty of Fundamental Problems of Technology, Wrocław University of Science and Technology, Wybrzeże Wyspiańskiego 27, 50-370 Wrocław, Poland}
\author{Anna Szerling}
\affiliation{Łukasiewicz Research Network -- Institute of Microelectronics and Photonics, al.~Lotników 32/46, 02-668 Warsaw, Poland}
\author{Tomasz Czyszanowski}
\affiliation{Photonics Group, Institute of Physics, Lodz University of Technology, ul.~Wólczańska 219, 90-924 Łódź, Poland}
\email{Corresponding author: e-mail: tomasz.czyszanowski@p.lodz.pl}


\begin{abstract}
Highly transparent and conductive electrodes operating in the infrared (IR) are critically needed for a broad range of technologies, including light-emitting diodes, lasers and photodetectors, which are key building blocks of infrared cameras, LiDARs, and thermal systems such as IR heaters. While transparent conductive electrodes (TCEs) have seen substantial progress in the visible spectrum, their performance in the IR remains limited due to increased absorption and reflection caused by the plasma resonance of free carriers in conductive materials. Here, we demonstrate a large-area TCE based on a metal-integrated monolithic high-contrast grating (metalMHCG) fabricated on a GaAs substrate. This structure acts as an effective antireflection coating, achieving near-unity transmission of unpolarized mid- to far-infrared (M-FIR) light. The metalMHCG exhibits 94\% transmission at a wavelength of \SI{7}{\micro \meter}, corresponding to 135\% relative to transmission through a flat GaAs–air interface, while maintaining an exceptionally low sheet resistance of \SI{2.8}{\OSq}. By simultaneously delivering excellent optical transparency and electrical conductivity, the metalMHCG establishes a new performance benchmark among M-FIR TCEs and provides a versatile platform for next-generation high-power optoelectronic devices.

\newline
\newline
 \end{abstract}
\maketitle


\section{Introduction}
Transparent conductive electrodes (TCEs) aim to balance two conflicting properties: high optical 
transparency and excellent electrical conductivity. Achieving high conductivity in TCEs requires a 
high concentration of free carriers in the TCE material, which inherently limits their transmittance. 
This fundamental trade-off between electrical conductivity and light transmission has been extensively 
studied in TCEs for visible spectrum (VIS) and near-infrared (NIR) applications, driving significant 
advancements in the field. As a result, TCEs presently play a crucial role in a wide range of 
optoelectronic devices, including sensors~\cite{Kim2023}, displays~\cite{Zhang2013}, light-emitting 
diodes~\cite{Hrong2017,Kwon2018,Jiang2018}, photovoltaics~\cite{Li2019,Zhang2014}, and flexible 
transparent electronics~\cite{Won2023}. 

Among the various methods for implementing TCEs, the use of indium tin oxide (ITO) is the most 
common~\cite{Granqvist2002}. Despite mass-scale production of ITO-based TCEs, its replacement is 
anticipated in the near future due to the scarcity of indium. Therefore, numerous approaches based on 
other transparent conductive oxides (TCO), graphene, thin metal plates or metal networks, and many 
more being investigated~\cite{Ellmer2012}.

In the mid-to-far infrared (M-FIR) spectral range, the primary application of TCEs is their 
integration with optoelectronic devices such as light emitting diodes (LEDs), photodetectors (PDs), 
lasers and cameras. Additionally, TCEs can serve as electromagnetic shields for M-FIR optoelectronic 
devices~\cite{Lai2023}, enhance the performance of transparent 
heaters~\cite{Zhang2017}, and improve the functionality of liquid crystal optical 
switches~\cite{Micallef2018}. The integration of TCEs with optoelectronic devices aims to increase the performance of the devices, but
poses fundamental challenges. The main challenge arises  from the resonance frequency of the free electron 
plasma in conductors (metals, nanocarbons, conductive oxides) which is located in the infrared 
spectrum~\cite{Chen2010,Ananthanarayanan2020}. Enhanced interaction of the electromagnetic field with 
free electrons results in high absorption and reflection in all conductive materials. In the case of 
conductive oxides that are transparent in VIS, such as ITO, $\mathrm{In_2O_3}$, $\textup{CuScO}_2$ and 
many more, lattice vibrations and impurity scattering further contribute to increased infrared 
absorption and a reduction in electrical conductivity, thereby limiting their practical use in 
infrared applications~\cite{Cui2023}.

In most TCE demonstrations, transmittance is typically defined as the percentage of light propagating 
through the TCE layer alone. However, when a TCE is deposited on a high-refractive-index substrate, 
reflection at the substrate surface becomes a non-negligible factor that further reduces overall 
transmission.  It seems inevitable that due to the intrinsic absorption of TCEs their implementation 
on the surface of the substrate reduces transmission compared to the case of a bare substrate–air 
interface. In this work, the transmission through the bare interface is referred to as the Fresnel 
limit ($T_{\fr}$), which can be calculated using the Fresnel formula:
 \begin{equation}
    T_{\fr}=1-\left(\frac{n_\text{s}-n_\text{a}}{n_\text{s}+n_\text{a}}\right)^2=\frac{4n_\text{s}n_\text{a}}{(n_\text{s}+n_\text{a})^2}
\end{equation}
where $n_{\textup{s}}$ and $n_{\textup{a}}$ are the refractive indices of the substrate and air, 
respectively, and the absorption of the substrate is assumed to be negligibly small. Surprassing the Fresnel limit 
remains a challenge for both VIS and IR TCEs, with the exception of Cu nanotrough networks operating 
in VIS, as demonstrated by C.~Ji, et al.~\cite{Ji2020}, which can surpass the Fresnel limit by 0.3\%. In the 
case of TCEs deposited on semiconductor devices, the Fresnel limit  is typically  less than 80\% in the 
case of wide bandgap semiconductors and less than 70\% in the case of narrow bandgap semiconductors. 

In our previous studies, we demonstrated that various configurations of monolithic high-contrast 
gratings integrated with metal (metalMHCG) can act as effective TCEs, enabling nearly 100\% transmission of 
infrared light polarized both along and perpendicularly~\cite{Ekielski2024,Tobing2021,Monvoisin2025} 
to the metalMHCG stripes. In this study, we develop the design and technology of metalMHCGs, presenting 
a method for achieving at the centimeter-scale surface near-perfect transparency of unpolarized (UPL) 
M-FIR light and record-low sheet resistance. In the proposed configuration, the metalMHCG enables the 
separation of carrier transport and radiation transmission channels, effectively minimizing strong 
infrared absorption and providing an efficient approach to enhancing transmission. Furthermore, 
the metalMHCG exhibits properties similar to those of anti-reflective layers, ensuring that its effective 
refractive index, as experienced by the interacting radiation, remains lower than that of the 
substrate, as in the case of conventional anti-reflective layers.

In the presented experimental demonstration, the metalMHCG consists of gold stripes embedded between 
periodically distributed GaAs stripes, structured on top of a GaAs substrate. This configuration 
achieves a transmission of 94\% for unpolarized light at a central wavelength of \SI{7}{\micro \meter} 
and results in a record-high relative transmittance of 135\% with respect to the Fresnel limit. 
Additionally, the transmission bandwidth exceeding the Fresnel limit is \SI{1.5}{\micro \meter}, 
corresponding to a relative bandwidth of 21\%.

The manuscript is structured as follows: Section~\ref{sec:conf} presents the configuration of the 
metalMHCG, focusing on the key design parameters and operational principles, supported by numerical 
simulations. Section~\ref{sec:fab} details the fabrication process of the metalMHCG. 
Sections~\ref{sec:opt} covers the optical and electrical characterization. Finally, 
Section~\ref{sec:dis} compares the properties of the metalMHCG with state-of-the-art TCEs and 
discusses its potential applications.
\section{Configuration and simulations}\label{sec:conf}
\begin{figure*}[ht]
\centering
\includegraphics[width=1.0\textwidth]{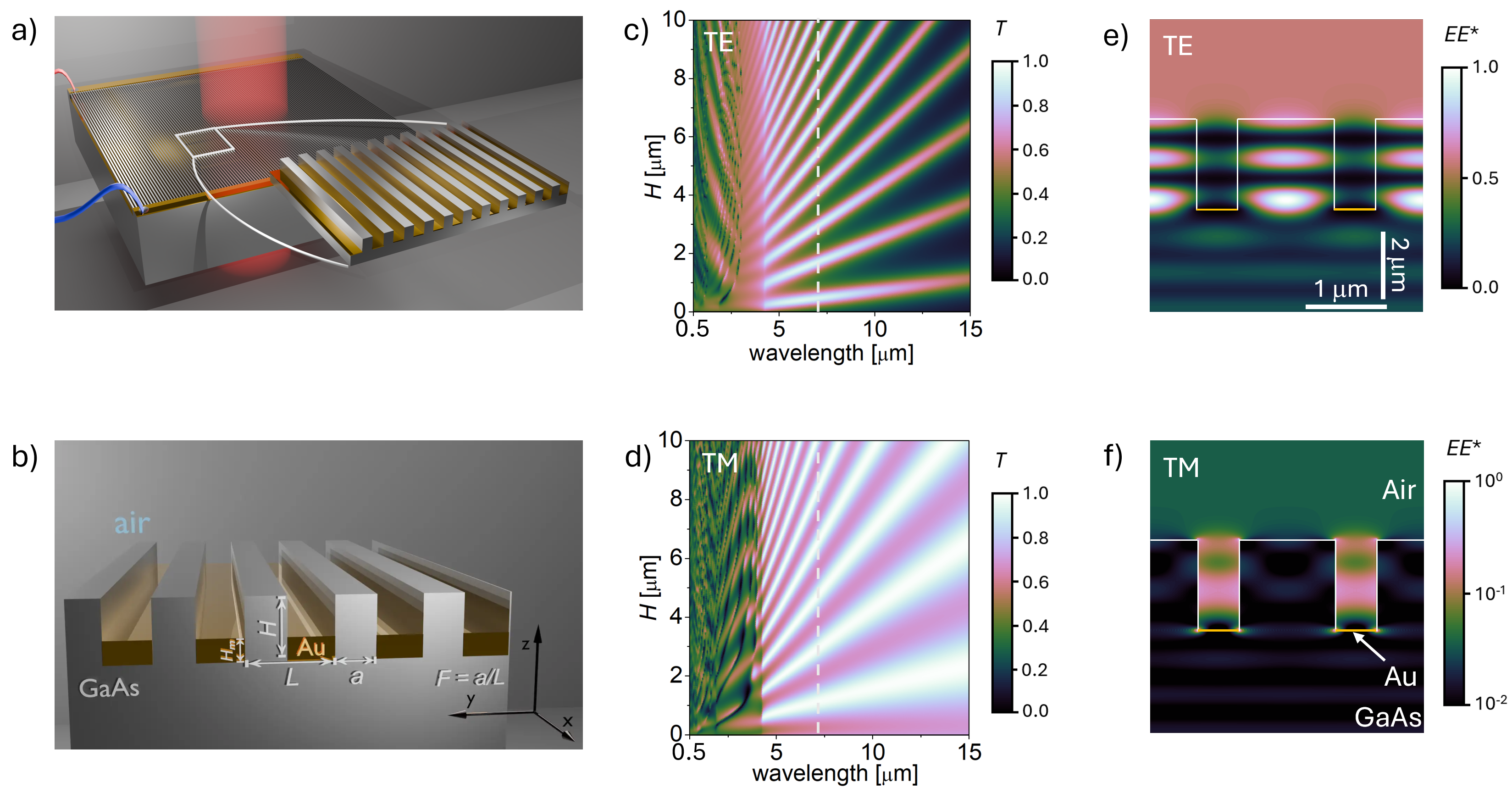}
\caption{\label{fig:1}a) Conceptual visualization of the metalMHCG composed of a one-dimensional grating 
on a GaAs wafer with gold stripes implemented in the grooves between the semiconductor stripes; b) 
Zoomed slice of the metalMHCG indicating the cross-section of the configuration, with geometrical 
definition of the grating parameters and the coordinate system. Calculated transmission ($T$) maps of 
the metalMHCG under normal incidence of c) TE and d) TM polarized light in the domain of the 
wavelength, and height of the semiconductor stripes ($H$) for $L=\SI{1.4}{\micro \meter}$, $F=0.74$, 
$H_\textup{m}=\SI{50}{\nano \meter}$. The white dashed lines in both figures indicate the wavelength 
of \SI{7}{\micro \meter}. Light intensity ($EE$*) distribution under normal incidence from the 
substrate side in the case of e) TE and f) TM polarization in the $yz$-plane of the metalMHCG cross-section.}
\end{figure*}

To demonstrate the transmission properties of metalMHCG using numerical methods, we consider the 
structure illustrated in Fig.~\ref{fig:1}, along with the Cartesian coordinate system adopted in 
analysis. The structure used in the calculations consists of a semi-infinite GaAs substrate, above 
which is a semi-infinite air superstrate. The surface of the GaAs in the $xy$-plane is 
patterned into an infinite one-dimensional grating consisting of parallel rectangular stripes along 
the $x$-direction, with alternating GaAs and gold stripes, both with a rectangular cross-section 
(Fig.~\ref{fig:1}a). The height of the GaAs stripes ($H$) exceeds that of the gold stripes 
($H_\textup{m}$). Additional parameters of the metalMHCG, also illustrated in Fig.~\ref{fig:1}b, 
include the period ($L$), the width of the semiconductor stripes ($a$), and their ratio, referred to 
as the fill factor ($F$). Light polarized along the $x$-direction (parallel to the stripes) is referred to as 
transverse-electric (TE) polarization while orthogonal polarization is referred to as transverse-magnetic (TM). 

In the numerical model, we utilize the plane wave admittance method~\cite{Dems2005} we used 
previously~\cite{Ekielski2024,Tobing2021,Monvoisin2025}, showing consistency with experimental 
results. The simulation considers the cross-section of the structure in the $yz$-plane, as illustrated 
in Fig.~\ref{fig:1}b. In the neglected $x$-direction, the solution is assumed to be a plane wave, 
corresponding to a plane wave incident normally on the metalMHCG surface. In the $y$-direction, we 
consider a single period of the grating with periodic boundary conditions, which extends the metalMHCG 
to infinity in this direction. We determine transmission for the case where light propagates 
perpendicularly to the metalMHCG plane from the substrate side to the air. The opposite propagation 
direction yields the same result.

In what follows, we consider an example illustrating an optimized structure that enables maximum 
transmission of unpolarized light for a wavelength of approximately $\lambda_\textup{max}=\SI{7}
{\micro \meter}$. The dimensions of the metalMHCG stripes, as defined in Fig.~\ref{fig:1}b, are 
$L=\SI{1.4}{\micro \meter}$, $F=0.74$, $H_\textup{m}=\SI{50}{\nano \meter}$. The dispersion of the 
refractive indices of GaAs ($n_\textup{GaAs}$) and Au ($n_\textup{Au}$) follow the experimental 
dependencies~\cite{Skauli2003,Olmon2012}. The optimization procedure for different refractive indices 
of the metalMHCG is detailed in~\cite{Sok2020}.

Figures~\ref{fig:1}c and~\ref{fig:1}d present numerically calculated transmission spectra as functions 
of the semiconductor stripe height ($H$) and the wavelength ($\lambda$) for TE and TM polarization, 
respectively. The maps exhibit two wavelength regions with noticeably different transmission 
properties. In the wavelength range $\lambda<n_\textup{GaAs}L\approx\SI{4.5}{\micro \meter}$, the 
transmission is the result of two modes interacting within the subwavelength grating, in what we refer to as the subwavelength region, as has been 
thoroughly discussed in~\cite{Chang-Hasnain2012}. In this 
region, only the zero-order diffraction of the grating can propagate in air, while more than one 
diffraction order can propagate in GaAs. Within this spectral range, the transmission through the 
metalMHCG can reach significant values, as indicated by the brighter areas in the transmission maps 
for both polarizations. However the spectral width of transmission above 70\% remains narrow. 

For $\lambda>n_\textup{GaAs}L$, the metalMHCG transmits and reflects only the zeroth diffraction order under normal incidence. In this spectral range, the metalMHCG 
exhibits properties typically observed in metastructures, characterized by an averaged interaction of 
both orthogonal polarizations with the structure~\cite{Pendry1999}. We will refer to this region as 
the deep subwavelength region, which is the primary focus of this article. In this region, high-transmission bands exist for both polarizations, which are significantly broader spectrally than the high-transmission regions present in the subwavelength region.

Analysis of Figs.~\ref{fig:1}c and~\ref{fig:1}d, indicates 
that both orthogonal polarizations display a typical etalon-like transmission dependence on the 
semiconductor stripe height ($H$). This behavior suggests that the metalMHCG functions as an etalon 
(Fabry-Perot (F-P) resonator) formed between the metalMHCG-air interface on one side and the metalMHCG–GaAs interface on the other, where the homogeneous GaAs substrate has a refractive index of 3.29 at a 
wavelength of \SI{7}{\micro\meter}~\cite{Czyszanowski2020}.

Based on the periodicity of transmission oscillations with respect to $H$ for both polarizations at a 
wavelength of $\lambda=\SI{7}{\micro\meter}$, 
one can determine the effective refractive indices of the metalMHCG, which are $n_\textup{effTE}=2.95$ and 
$n_\textup{neffTM}=2.11$ for TE and TM polarizations, respectively. The different periodicities for 
the two polarizations and hence different effective refractive indices result from different spatial 
field distributions of TE and TM polarizations within the metalMHCG. Figure~\ref{fig:1}e 
and~\ref{fig:1}f indicate that TE polarization is mainly confined within the semiconductor stripes, 
whereas TM polarization is confined in the air gaps between the stripes. Additionally, the optical field of both 
polarizations penetrates the metal stripes only to a limited extent. Plasmonic effects at the GaAs-gold 
interface for TM polarization are almost entirely suppressed compared to similar structures composed 
solely of parallel metal stripes~\cite{Porto1999}. Interestingly, in the MHCG without metal, it is impossible to achieve such high transmission as can be obtained in the metalMHCG (for unpolarized light)~\cite{Sok2020}. 

The periodic behaviour of transmission with respect to the height of the metalMHCG stripes, which differs 
for each polarization, enables an optimal $H$ to be found at which both polarizations achieve nearly 100\% 
transmission, resulting in almost complete transmission of unpolarized light. This condition can be 
satisfied with relatively small $H$ by tuning $F$, which affects the effective refractive indices of both polarizations.

\begin{figure*}[ht]
\centering
\includegraphics[width=1.0\textwidth]{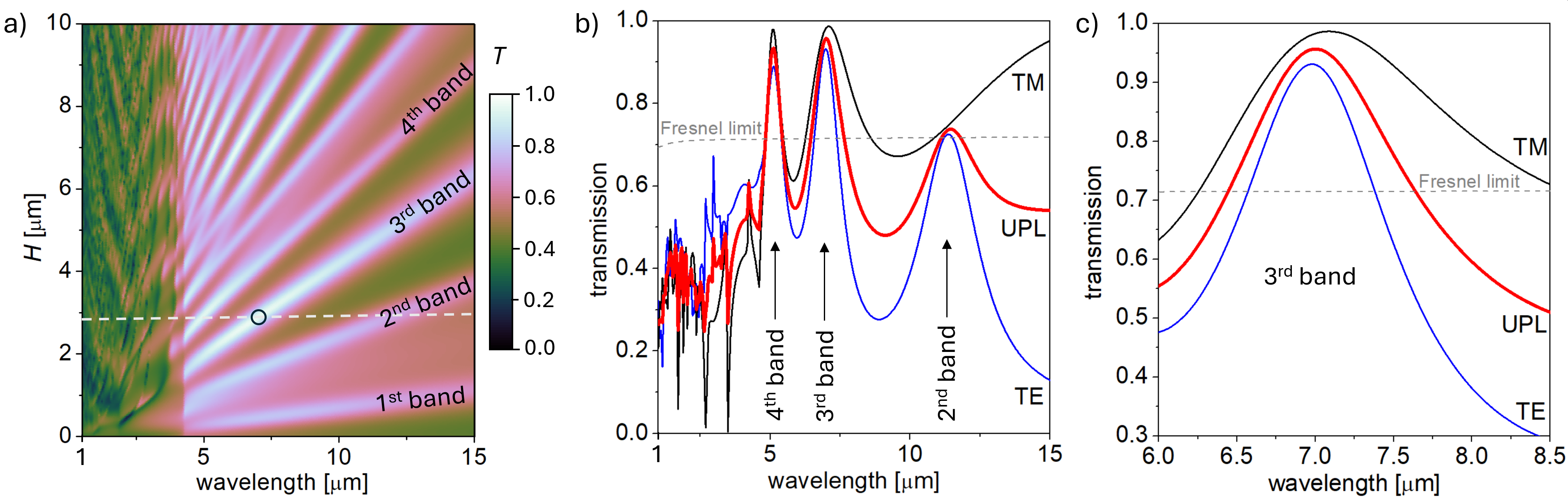}
\caption{\label{fig:2}a) Calculated transmission ($T$) map of the metalMHCG under normal incidence of 
unpolarized light in the domain of the wavelength, and height of the semiconductor stripes ($H$) for 
$L=\SI{1.4}{\micro \meter}$, $F=0.74$, $H_\textup{m}=\SI{50}{\nano \meter}$, where the white dashed line 
indicates $H=\SI{2.79}{\micro \meter}$ and black circle indicates the global maximum of the transmission at 
the wavelength of \SI{7}{\micro \meter}; b) calculated transmission spectra of TE (blue) and TM (black) 
polarizations as well as unpolarized (red) light along the white dashed line from a); c) zoom of the spectrum in 
the proximity of transmission maximum. The Fresnel limit indicates transmission through the plane interface 
between GaAs and air in b) and c).}
\end{figure*}

Figure \ref{fig:2}a presents the transmission map of unpolarized light through the metalMHCG. The high-
transmission bands, which form regular regions in the ($H, \lambda$) domain for polarized light (see 
Fig. \ref{fig:1}c, \ref{fig:1}d), transform to a more complex pattern for unpolarized light due to the different effective refractive indices of the two polarizations. Consequently, the transmission of 
unpolarized light exhibits non-periodic  behavior. The successive high-transmission bands of 
unpolarized light, occurring with increasing $H$, exhibit transmission maxima of different values.

Figure \ref{fig:2}b presents the transmission spectrum of unpolarized light (UPL) for the optimized 
configuration defined by the parameters $L=\SI{1.431}{\micro\meter}$, $F=0.734$, $H=\SI{2.787}
{\micro\meter}$, and $H_\textup{m}=\SI{50}{\nano\meter}$, in which the second, third, and fourth 
transmission bands are indicated. Within the third transmission band, the global transmission maximum 
reaches $T_\textup{max}^{(3)}=95.7\%$, while transmission exceeding the Fresnel threshold 
($\Delta\lambda_\fr^{(3)}$) spans over \SI{1.2}{\micro\meter}, corresponding to a relative spectral width 
($\Delta\lambda_\fr^{(3)}/\lambda_\textup{max}^{(3)}$) of more than 17.2\%. In the fourth transmission band, these 
values are $T_\textup{max}^{(4)}=93.4\%$,  $\Delta\lambda_\fr^{(4)}/\lambda_\textup{max}^{(4)}=12.5\%$, and in 
the second transmission band $T_\textup{max}^{(2)}=73.7\%$,  $\Delta\lambda_\fr^{(2)}/\lambda_\textup{max}^{(2)}=5.0\%$.

The maximum transmission is achieved with slightly different peak values for TE and TM polarizations, 
as shown in Fig. \ref{fig:2}c. While equalizing the transmission of both polarizations is possible 
by adjusting the metalMHCG parameters, it comes at the cost of a 3\% reduction in total transmission 
in the analyzed case. A more comprehensive numerical analysis of the optical properties, including the 
angular dependence of transmission and impact of metalMHCG composition on transmission, is presented in \cite{Sok2020}. 

The metalMHCG reveals superb electrical properties, as the volume of the metal in metalMHCG is 
significantly larger than in any other TCE used in any spectral range. Considering the geometric 
parameters of the gold stripes ($H_\textup{m}=\SI{50}{\nano\meter}$, $L-a=\SI{364}{\nano\meter}$) and 
assuming a bulk gold resistivity of ($2.44\cdot10^{-8}$ \SI{}{\ohm \meter}), the optimized metalMHCG 
exhibits a sheet resistance of \SI{2}{\OSq}.

\section{Fabrication}\label{sec:fab}
\begin{figure*}[ht]
\centering
\includegraphics[width=1.0\textwidth]{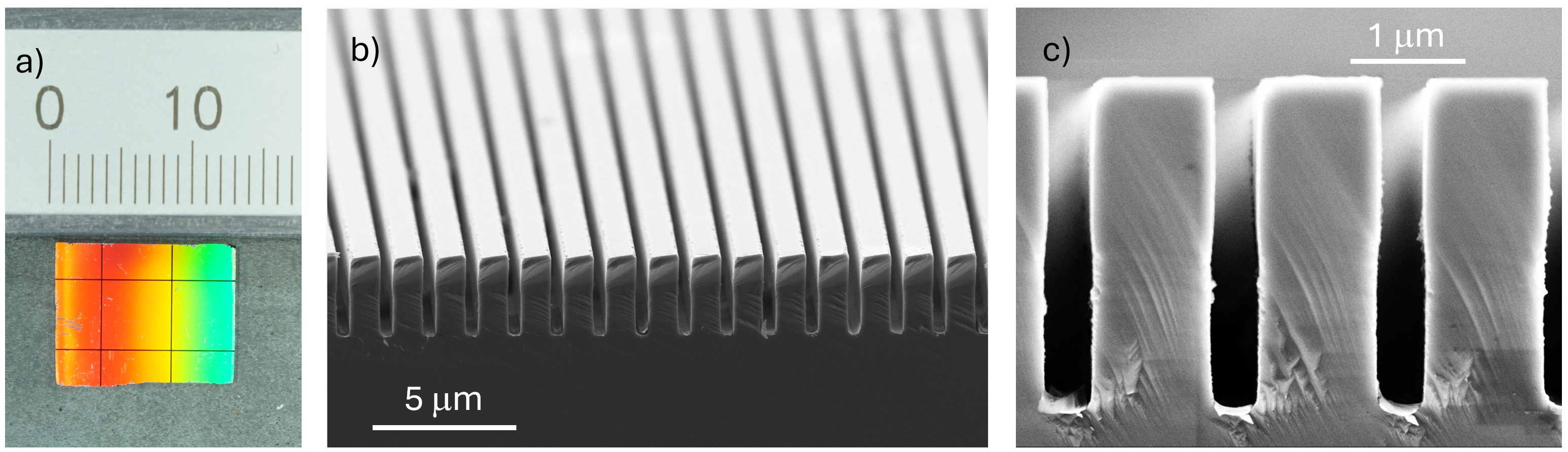}
\caption{\label{fig:3}a) Visible light image of the sample, black lines are spacers between metalMHCG patches, b) and c) cross-section scanning electron microscope (SEM) images of metalMHCG in various magnifications. Metal stripes are visible at bottom of the grooves in~c).}
\end{figure*}

The fabricated metalMHCG covers more than 1\,$\textup{cm}^2$ of a GaAs wafer, in the form of nine patches. The central square-shape patch has a side length of $5\times 5$\,mm (Fig.~\ref{fig:3}a). The nominal geometric 
parameters of the realized structure are the same as those of the metalMHCG structure used in the 
numerical analysis. An anti-reflection coating is not deposited on the opposite wafer surface, due to 
the significant infrared absorption of the dielectric materials.

The realization of the GaAs-based metalMHCG with gold stripes involves a combination of plasma-enhanced 
chemical vapor deposition (PECVD), electron beam lithography (EBL), inductively coupled plasma-
reactive ion etching (ICP-RIE), and e-beam physical vapor deposition (EBPVD). The two main 
technological challenges are the fabrication of a GaAs grating with parallel grooves with an aspect 
ratio of depth to width of 7.7 and the accurate placement of metal stripes at the bottom of these 
grooves (see Fig.~\ref{fig:3}b, c).

Our high aspect ratio metalMHCG TCE was created by transferring the pattern of the grating generated 
in the EBL process from resist to GaAs substrate through three layers of hard masks consisting of $
\mathrm{SiO_2/Cr/SiO_2}$ in the ICP-RIE plasma etching procedure. The pattern was transferred to the 
substrate using $\mathrm{BCl_3/N}_2$ in short cycles of etching and cooling, achieving 
smooth, slightly concave grating stripes profile. This profile prevents the deposition of metal from deposition on side walls of the 
groves, which could potentially contribute to  absorption. This shape allowed for precise gold 
deposition on the bottom of the grooves and easy removal of the leftover hard mask in the buffered hydrofluoric acid solution.

The actual cross-section shape and dimensions of the processed metalMHCG were determined based on SEM 
images: $L=\SI{1.47}{\micro\meter}$ (with standard deviation of \SI{0.01}{\micro\meter}), $F=\SI{0.747}
{\micro\meter}$, $H=\SI{2.89}{\micro\meter}$ (standard deviation of \SI{0.04}{\micro\meter}), 
$H_\textup{m}=\SI{51}{\nano\meter}$ (standard deviation of 2\,nm).

\section{Optical and electrical properties}\label{sec:opt}
Transmission measurements were conducted using a Vertex 80v vacuum Fourier Transform Infrared 
spectrometer (FTIR) from Bruker. The light was generated by a polychromatic source (either a halogen lamp
or a glow bar, depending on the spectral range) and focused by an optical system with parabolic mirrors 
onto the metalMHCG at a normal incident angle, creating a roughly 1-mm diameter spot entirely 
contained within the $5\times 5$\,mm metalMHCG area. The spot size causes the transmission measurement 
to be averaged over more than 700 periods of the metalMHCG, which is therefore subject to statistical 
error due to fabrication imperfections, such as the waviness of the semiconductor stripes, inhomogeneity, 
and discontinuities in the gold stripes. The light intensity was measured using a HgCdTe (MCT) 
liquid-nitrogen cooled detector. As references, transmission spectra were recorded for an empty 
chamber and for a piece of the same wafer without a metalMHCG. This approach enabled the measurement 
of light transmission while eliminating the influence of potential absorption and scattering in the 
wafer, as well as Fresnel reflection from the wafer’s opposite surface. 

\begin{figure*}[ht]
\centering
\includegraphics[width=1.0\textwidth]{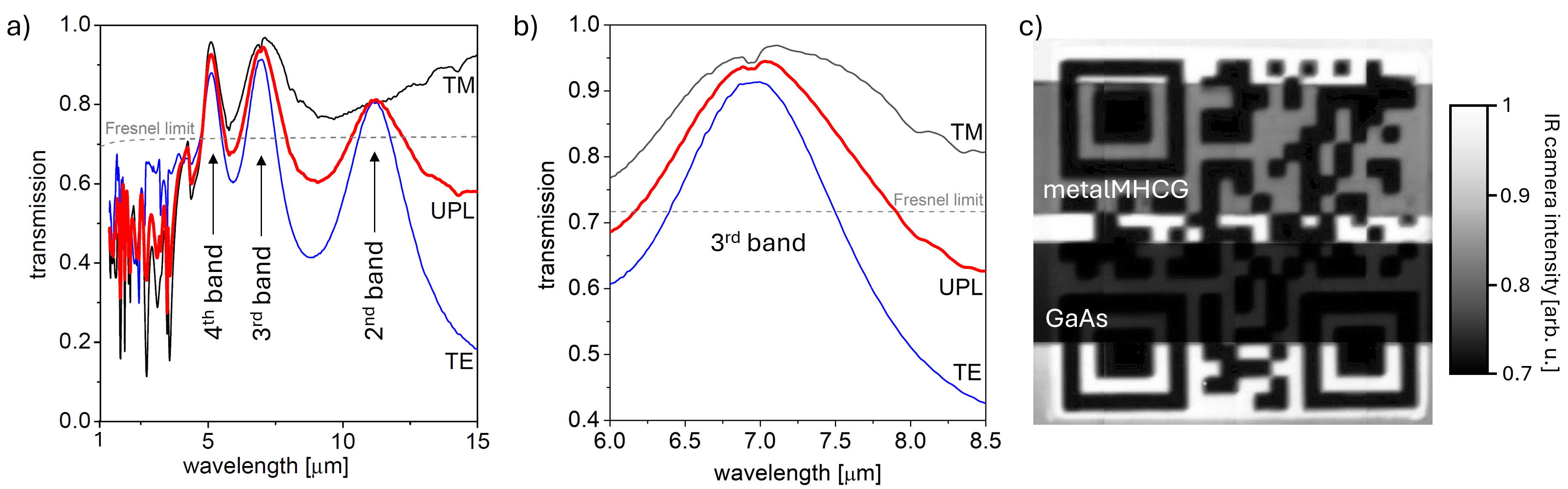}
\caption{\label{fig:4}a) Experimental transmission spectra of the metalMHCG in the case of TE (blue) and TM 
(black) polarizations as well as unpolarized (red) light; b)~zoom of the experimental spectrum in the 
proximity of the transmission maximum in the third band. Fresnel limit indicates transmission 
through plane interface between GaAs and air; c)~infrared image taken in the wavelength range of 4.8–\SI{5.0}{\micro\meter} (fourth band), showing a QR code heated to \SI{80}{\degreeCelsius} in the background and two samples from the same GaAs wafer: metalMHCG on the GaAs substrate 
(metalMHCG) and bare GaAs substrate (GaAs) placed above the QR code.}
\end{figure*}

The experimentally measured transmission spectrum (Fig.~\ref{fig:4}) closely aligns with the 
numerically computed spectrum. In the deep 
subwavelength regime, three distinct transmission maxima for unpolarized light are observed, 
corresponding to the second, third, and fourth transmission bands, with peak wavelengths at \SI{5.1}
{\micro\meter}, \SI{7.1}{\micro\meter}, and \SI{11.2}{\micro\meter} and maximal transmission of 92.5\%, 
94\%, and 81\%, respectively. Within the subwavelength range ($\lambda<\SI{4.5}{\micro\meter}$), 
multiple narrow spectral regions exhibit pronounced reflection and transmission features, which, 
according to numerical simulations, can exceed 90\% within a very narrow bandwidth. However, in the 
specific configuration studied here, the transmission maxima in this range remain below the Fresnel 
limit. In the long-wavelength region near \SI{15}{\micro\meter}, the experimentally measured 
transmission reaches approximately 60\%, which aligns well with numerical simulations. According to 
these calculations, at even longer wavelengths beyond the experimental range the transmission remains 
relatively stable, at around 50--60\%.

 The primary goal in designing the metalMHCG structure was to achieve the highest possible 
 transmission, positioned arbitrarily within the M-FIR range, with a spectral bandwidth comparable to 
 the emission or absorption linewidths of optoelectronic devices operating in this region. 
 Consequently, our analysis focuses on the third transmission band, which aligns with these objectives 
 and exhibits a global transmission maximum of 94\% for the given configuration.  The experimentally 
 determined maximum transmission is only slightly lower than the theoretical prediction of 95.7\%. For 
 TM polarization, the transmission reaches 96.9\%, surpassing the TE polarization transmission, which 
 reaches 91.3\%. The peak unpolarized light transmission, normalized to the Fresnel level (relative 
 transmission), reaches 135\% for the analyzed metalMHCG---significantly exceeding previously reported 
 maximum values~\cite{Ji2020,Ekielski2024}. The spectral width of unpolarized light transmission 
 exceeding the Fresnel limit is approximately \SI{1.5}{\micro\meter}, corresponding to a relative 
 spectral bandwidth of 21\% with respect to the peak wavelength. The broader transmission spectrum 
 compared to the numerical results presented in Section~\ref{sec:conf} is attributed to the non-
 rectangular cross-section of the stripes, which enhances the transmission bandwidth.  

To evaluate the transmission properties of the metalMHCG, we performed infrared imaging using an InSb-
cooled infrared camera equipped with a 1/4 inch extension ring to enhance 
magnification~\cite{Kaluza01012024} and a 4.8–\SI{5.0}{\micro\meter} bandpass optical filter, selected 
to match the spectral region of the fourth transmission band of the metalMHCG. 
Figure~\ref{fig:4}c shows a thermographic image of a QR code patterned with FR4 laminate on copper 
cladding at a temperature of 80\textdegree C. The horizontal, darker, rectangle-shaped regions 
correspond to areas imaged through a bare GaAs wafer (bottom sample) and through the GaAs wafer with 
the metalMHCG structure deposited on one of its interfaces (top sample). In both samples, light is 
reflected by the flat bottom GaAs–air interface. The difference in brightness indicates higher 
transmission through the opposite interface with the metalMHCG (top sample) compared to the flat GaAs–
air interface (bottom sample).

To accurately evaluate the sheet resistance of the metalMHCG structure, an indirect approach is 
required. Due to the inherently low resistance of the fabricated large-area metalMHCG, its total 
resistance is significantly lower than the parasitic resistance introduced by the electrical contacts 
and probing system, as confirmed by short-circuit configuration measurements. Since the total 
resistance of the metalMHCG is primarily governed by its metallic components, we evaluate the sheet 
resistance using dedicated test structures. These structures consist of a reduced number of parallel 
gold stripes, allowing the overall resistance to reach a level that is measurable with standard 
laboratory equipment while minimizing errors associated with the parasitic resistance of the measurement setup.

The test structures were fabricated on a non-conductive silicon substrate and include configurations 
with 4, 6, 8, and 12 gold wires, connected in parallel, each 2\,mm in length and with nominal 
dimensions of 50\,nm in height and 365\,nm in width. These dimensions are identical to those used in the metalMHCG. 

 The current–voltage measurements exhibited linear behavior, allowing the electrical conductivity of 
 a single gold stripe to be determined as approximately 0.25\,mS. Based on the stripe dimensions and the metalMHCG period, the corresponding 
 sheet resistance is estimated to be around \SI{2.8}{\OSq}. This value is only about 40\% higher than 
 the theoretical sheet resistance calculated using the bulk conductivity of gold, which does not 
 account for size effects in thin metallic wires or the presence of impurities. The electrical 
 properties of the fabricated metalMHCG are consistent with those reported in a previously 
 demonstrated metalMHCG configuration designed for polarized light ~\cite{Ekielski2024}.

\section{Discussion}\label{sec:dis}
\begin{figure*}[ht]
\centering
\includegraphics[width=0.7\textwidth]{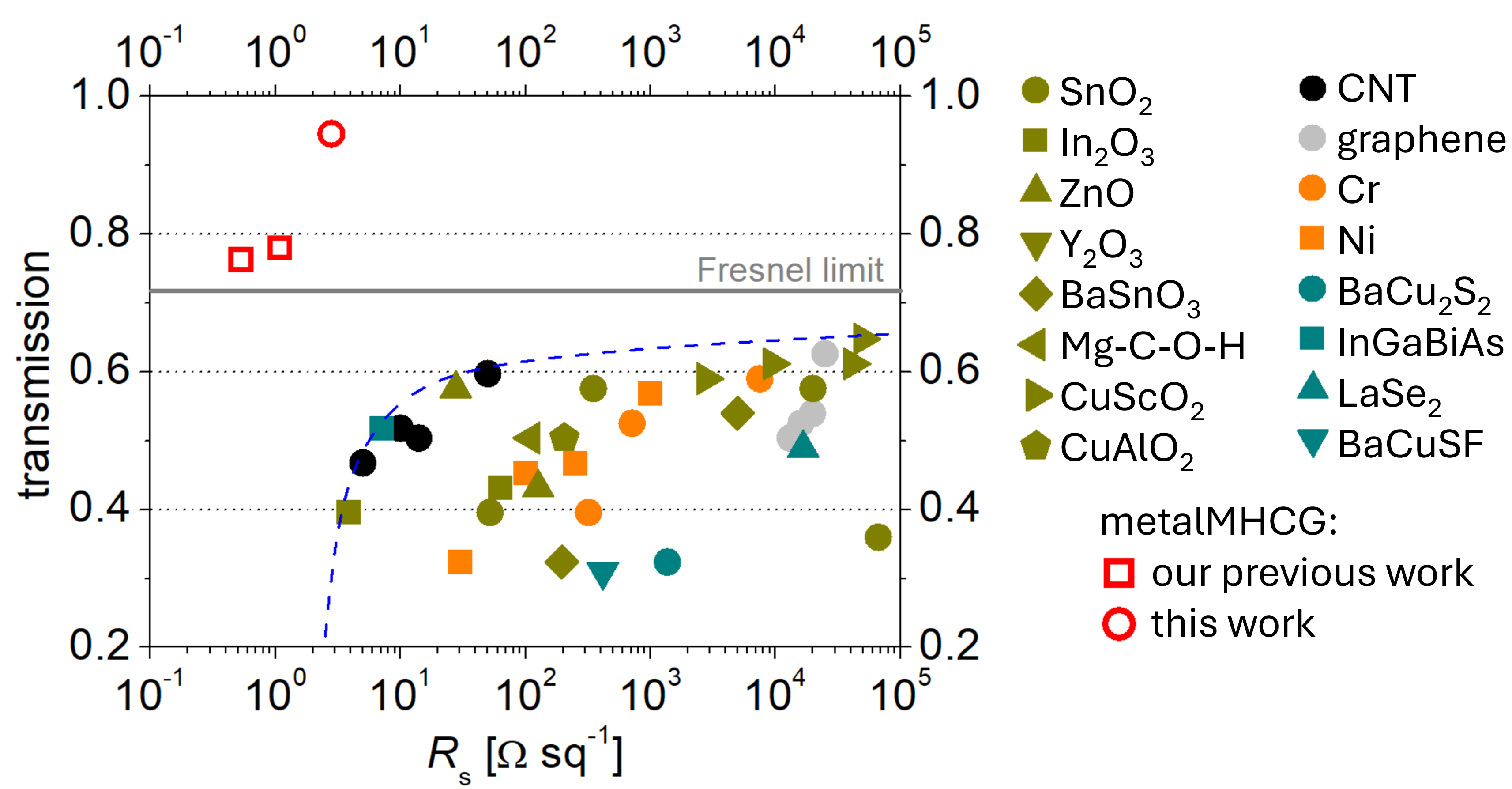}
\caption{\label{fig:5}Optical maximal transmissions at M-FIR as a function of sheet resistance for 
TCEs based on various material approaches: ($\textup{SnO}_2$~\cite{Xu2021,Fukumoto2022,Geraldo2003,Park2020}, $\mathrm{In_2O_3}$~\cite{Guo2019,Du2016}, 
ZnO~\cite{Cheng2020,Fernandes2013}, $\mathrm{Y_2O_3}$~\cite{Yang2013}, $\textup{BaSnO}_3$~\cite{Seo2014}, Mg-C-O-H~\cite{Guo2019Yang}, $\textup{CuScO}_2$~\cite{Chuai2015,Chuai2016}, $\textup{CuAlO}_2$~\cite{Johnson2001}, carbon nanotube (CNT)~\cite{Ma2007}, graphene~\cite{Han2016}, 
Cr~\cite{Ghosh2009}, Ni~\cite{Ghosh2009}, $\mathrm{BaCu_2S_2}$~\cite{Park2002}, 
InGaBiAs~\cite{Zhong2013}, $\textup{LaSe}_2$~\cite{Gao2021}, BaCuSF~\cite{Frantz2013}, metalMHCG our 
previous work~\cite{Ekielski2024}). The reported transmittance has been recalculated to represent the transmission through the TCE deposited on a GaAs substrate, the grey dashed line presents transmission Fresnel limit at \SI{7}{\micro\meter}. The red circle represents unpolarized transmission near \SI{7}{\micro\meter} for the metalMHCG presented in this work.}
\end{figure*}

We compare the properties of the metalMHCG with other TCEs operating in the mid-infrared range by 
evaluating their optical and electrical characteristics, as reported in 
references~\cite{Xu2021,Fukumoto2022,Geraldo2003,Park2020,Guo2019,Du2016,Cheng2020,Fernandes2013,Yang2013,Seo2014,Guo2019Yang,Chuai2015,Chuai2016,Johnson2001,Ma2007,Han2016,Ghosh2009,Park2002,Zhong2013,Gao2021,Frantz2013}. In most cases, the transmittance or relative transmission of these TCEs was assessed 
in configurations where they were implemented on low-refractive-index substrates such as glass or 
polymers. For the purpose of a consistent comparison, the transmission values of the TCEs were 
recalculated assuming their integration on a GaAs substrate with a representative refractive index 
of~3.29, for which the corresponding Fresnel limit is 71.5\%. Figure~\ref{fig:5} presents various TCEs 
as data points in the space defined by maximum transmission and sheet resistance. The distribution of 
the points corresponding to the most efficient configurations, excluding the red-marked data points 
representing our previous and current work, follows a trend indicated by the blue dashed line. This 
trend shows that achieving transmission exceeding 60\% (corresponding to a transmission relative to 
the Fresnel limit $T_\textup{R}>80\%$) is typically possible only for devices with a sheet resistance 
greater than \SI{50}{\OSq}. A notable example is a 200\,nm-thick $\textup{CuSCO}_2$ 
layer~\cite{Chuai2016}, which exhibits the highest reported transmission of 64\% ($T_\textup{R}=90\%$) 
and sheet resistance of $50\,\mathrm{k\Omega/sq^{-1}}$. At the opposite end of the trade-off line, a 
\SI{1.3}{\micro\meter}-thick $\mathrm{In_2O_3}$ layer~\cite{Guo2019} demonstrates the lowest reported 
sheet resistance of \SI{3.8}{\OSq}, accompanied by a transmission below 40\% ($T_\textup{R}=55\%$). 
Among the considered TCEs, a 100\,nm-thick carbon nanotube (CNT)-based layer~\cite{Ma2007} offers an 
attractive balance between optical and electrical performance, with a transmission of 55\% 
($T_\textup{R}=83\%$) and a sheet resistance of~\SI{50}{\OSq}.

MetalMHCGs are the only TCEs that surpass the Fresnel limit and are positioned significantly above the 
trend line. Our previous work~\cite{Ekielski2024}, which demonstrated a metalMHCG operating in the 
first transmission band (see~\ref{fig:5}a) and optimized for efficient transmission of polarized 
radiation, is marked in the figure as a red hollow square. This configuration achieved a record-low 
sheet resistance of \SI{0.5}{\OSq} and an unpolarized light transmission of 73\% ($T_\textup{R}=106\%$).
The TCE presented in this work, indicated by the red circle, outperforms all non-metalMHCG TCEs in both 
parameters simultanously, offering 1.7~times higher transmission and 30\% lower sheet resistance than 
any best-performing TCEs in either category. While the transmission of metalMHCG exhibits greater 
spectral variability than that of conventional TCEs---due to its reliance on a low quality-factor 
resonance---its spectral width is typically broader than the emission spectra of IR light sources such 
as LEDs or lasers, and it can be arbitrarily positioned within the M-FIR spectrum by design.

TCEs operating in the M-FIR face at least two key challenges compared to those designed for the 
visible and NIR ranges. The proximity to the plasma frequency of free electrons leads to significantly 
higher absorption, and integration with emitters such as quantum cascade lasers or interband cascade 
LEDs and lasers imposes the need to sustain very high current densities without excessive power 
dissipation or performance degradation. These devices require the highest current densities among all 
optoelectronic systems. As a result, M-FIR TCEs must achieve a sheet resistance below \SI{10}{\OSq}, 
an order of magnitude lower than their visible-range counterparts~\cite{Cao2013}.

While the metalMHCG demonstrated in this work does not inherently function as an electrical contact, 
only minor design modifications are required to achieve efficient electrical conductivity between the 
metal and the semiconductor, without causing any significant degradation in optical 
transmission~\cite{Monvoisin2025}. Nevertheless, its record-low sheet resistance, combined with 
exceptionally high transmittance, positions metalMHCGs as a leading candidate as TCE for high-power 
optoelectronic applications. In photodetectors employed in optical communication, imaging 
systems, laser guidance, and biosensing, such highly transparent electrodes enhance carrier collection 
efficiency and reduce response times by enabling faster carrier transport from the active region to 
the electrode. Additionally, metalMHCGs are suitable for shielding applications in aviation, military, and medical fields. They can protect 
sensitive electronics from electromagnetic fields at wavelengths longer than IR, due to their dense metal 
distribution, while preserving image clarity in the infrared range due to their deep-subwavelength geometry. 
Finally, metalMHCGs are well-suited for use as transparent heaters in energy-efficient heating applications 
and as infrared liquid crystal optical switches for advanced light modulation technologies. 

The proposed metalMHCG exceeded the Fresnel limit, which had never been significantly surpassed 
before, and approached total transmission, clearly outperforming all previously proposed solutions and setting a new performance benchmark.
\input{arxiv.bbl}
\subsection*{Acknowledgments}
MMo acknowledges the support from the Polish National Science Center, grant OPUS 2019/33/B/ST7/02591. AS and ME acknowledge the support by the statutory funds of the Łukasiewicz Research Network – Institute of Microelectronics and Photonics. This work has been completed while KB, WG and MZ were the Doctoral Candidate in the Interdisciplinary Doctoral School at the Lodz University of Technology, Poland.
\subsection*{Author contributions}TC conceived this research project. KB, ME, OS, MZ and AS fabricated the metalMHCGs. WG and TC performed the numerical modelling. TS, MR and MMo conducted the optical measurements. MMa, MG, and MW performed electrical measurements. MK, TS, ME and MW visualised the samples. TC and AS directed the project. TC wrote the manuscript with input from all authors. All authors discussed the results and contributed to the manuscript. 

\end{document}

%% file: arxiv.bbl
%